\documentclass[12pt]{article}
\usepackage{times}
\usepackage{geometry}
\usepackage[utf8]{inputenc}
\usepackage{enumitem,amssymb}
\usepackage{ragged2e}
\newlist{thematic}{itemize}{8}
\setlist[thematic]{label=$\square$}
\usepackage{pifont}

\usepackage{epsfig,wrapfig,paralist}
\usepackage[numbers]{natbib}

\newcommand{\ciii}{\hbox{C\,{\sc iii}]}}

\newcommand{\civ}{\hbox{C\,{\sc iv}}}

\newcommand{\nat}{Nature}
\newcommand{\apj}{ApJ}
\newcommand{\aj}{AJ}
\newcommand{\aap}{A\&A}
\newcommand{\mnras}{MNRAS}
\newcommand{\apjl}{ApJL}

\begin{document}
\begin{raggedright}
\setlength{\parindent}{1cm}
\huge
\begin{centering} Astro2020 Science White Paper \linebreak

\noindent Unveiling the Phase Transition of the Universe During the Reionization Epoch with Ly$\alpha$ \linebreak
\end{centering}
\normalsize

\noindent \textbf{Thematic Areas:} \hspace*{60pt} $\square$ Planetary Systems \hspace*{10pt} $\square$ Star and Planet Formation \hspace*{20pt}\linebreak
$\square$ Formation and Evolution of Compact Objects \hspace*{31pt} \makebox[0pt][l]{$\square$}\raisebox{.15ex}{\hspace{0.1em}$\checkmark$} Cosmology and Fundamental Physics \linebreak
  $\square$  Stars and Stellar Evolution \hspace*{1pt} $\square$ Resolved Stellar Populations and their Environments \hspace*{40pt} \linebreak
  \makebox[0pt][l]{$\square$}\raisebox{.15ex}{\hspace{0.1em}$\checkmark$} Galaxy Evolution   \hspace*{45pt} $\square$             Multi-Messenger Astronomy and Astrophysics \hspace*{65pt} \linebreak
  
\noindent \textbf{Principal Author:}

\noindent Name:	Steven L. Finkelstein
 \linebreak						
Institution:  The University of Texas at Austin
 \linebreak
Email: stevenf@astro.as.utexas.edu 
 \linebreak
Phone:  (512) 471-1483
 \linebreak
 
\noindent \textbf{Co-authors:} 
Marusa Bradac (UC Davis),
Caitlin Casey (UT Austin),
Mark Dickinson (NOAO),
Ryan Endsley (Arizona),
Steven Furlanetto (UCLA), 
Nimish Hathi (STScI),
Taylor Hutchison (Texas A\&M),
Intae Jung (UT Austin),
Jeyhan Kartaltepe (RIT),
Anton M. Koekemoer (STScI), 
Rebecca L. Larson (UT Austin),
Charlotte Mason (Harvard/Smithsonian CfA),
Casey Papovich (Texas A\&M),
Swara Ravindranath (STScI),
Jane Rigby (NASA GSFC),
Dan Stark (Arizona),
Isak Wold (NASA GSFC)
  \linebreak

\noindent \textbf{Abstract  (optional):}
The epoch of reionization (6 $\lesssim z \lesssim$ 10) marks the period in our universe when the first large galaxies grew to fruition, and began to affect the universe around them.  Massive stars, and potentially accreting supermassive black holes, filled the universe with ionizing radiation, burning off the haze of neutral gas that had filled the intergalactic medium (IGM) since recombination ($z \sim$ 1000).  The evolution of this process constrains key properties of these earliest luminous sources, thus observationally constraining reionization is a key science goal for the next decade.  The measurement of Ly$\alpha$ emission from photometrically-identified galaxies is a highly constraining probe of reionization, as a neutral IGM will resonantly scatter these photons, reducing detectability.  While significant work has been done with 8--10m telescopes, \emph{\textbf{these observations require extremely large telescopes}} (ELTs) – the flux limits available from today’s 10m class telescopes are sufficient for only the brightest known galaxies ($m <$ 26). Ultra-deep surveys with the Giant Magellan Telescope (GMT) and Thirty Meter Telescope (TMT) will be capable of detecting Ly$\alpha$ emission from galaxies 2-3 magnitudes fainter than today's deepest surveys.  Wide-field fiber spectroscopy on the GMT combined with narrow-field AO-assisted slit spectroscopy on the TMT will be able to probe the expected size of ionized bubbles throughout the epoch of reionization, following up $\sim$degree scale deep imaging surveys with the \emph{Wide Field Infrared Space Telescope}.  These data will provide the first resolved Ly$\alpha$-based maps of the ionized intergalactic medium throughout the epoch of reionization, constraining models of both the temporal and spatial evolution of this phase change. 
\pagebreak 

\large
\noindent \textbf{Reionization: The Last Major Phase Transition}  
\normalsize
\vspace{1mm}
\end{raggedright}

\noindent At the time photons from the surface of last scattering began their journey to our
telescopes, the universe had cooled enough for hydrogen
gas to become neutral.  This was the state for the next few hundred
million years --- the so-called ``dark ages''.  Eventually, baryonic
matter cooled and condensed in dark matter halos to form the first
stars, galaxies, and growing supermassive black holes.  These objects emitted hydrogen ionizing photons, burning away the cosmic haze of neutral gas.  This reionization of the intergalactic medium (IGM) was the last major phase change in the universe, and is inextricably linked with the onset of
the first luminous sources in our history.  Therefore, understanding how reionization proceeds, both temporally and spatially, provides key constraints on the nature of these first luminous sources in the universe.
\vspace{-6mm} 

\section{\large Present-Day Constraints on Reionization}
\vspace{-4mm} 

Improving our knowledge of the epoch of reionization (EoR; hereafter 6 $\lesssim z \lesssim$ 10) was a key focus of the New Worlds New Horizons decadal survey.  At the end of this decade, while we have learned much about galaxies in this epoch (and their potential contribution to the ionizing background) from deep near-infrared imaging
surveys with the {\it Hubble Space Telescope} ({\it HST}) and wide-field ground-based surveys \citep[e.g.,][]{finkelstein15,bouwens15,mclure13,livermore17,atek18,bowler14,bowler15}, our
constraints on the evolution of the reionization process are presently poor. 
However, we do have a reasonable idea when reionization completed.  Observations of the Gunn-Peterson trough in high-redshift quasars measure the neutral fraction ($Q_{HI}$) to be $\ll$1\% at
$z \approx$ 6 \citep[e.g.,][]{fan06}.  While this result is dependant on assumptions about the intrinsic shape of the quasar's rest-ultraviolet (UV) spectrum, model-independent constraints agree that the neutral fraction is low at $z \sim$ 6, though constraints are looser ($Q_{HI} <$ 0.06 $\pm$ 0.05 at $z \sim$ 5.9; \cite{mcgreer15}).  Significant neutral patches may be necessary to explain the most opaque stretches of the Ly$\alpha$ forest at $z=5.5-6$; in
particular, the observed $\sim 110h^{-1}$ comoving Mpc Ly$\alpha$ trough \cite{becker15,kulkarni18}.  However, there is little evidence that reionization competed at later times than $z \sim$ 5.5.

While knowledge of the completion redshift is key, understanding early structure formation requires constraints on the \emph{entire process} of reionization.  Observations of the optical depth due to electron scattering from the cosmic microwave background imply that the average redshift where a region of the universe undergoes reionization is at $z \sim$ 7--8 \citep{planck16}.
Observations of the few known $z \geq$ 7 quasars \citep{mortlock11,bolton11,greig17b,banados18} find $Q_{HI}$ $>$ 10\% at $z =$ 7, and possibly as high as 50\% at $z =$ 7.5. However, these measurements again depend on our ability to model the intrinsic quasar spectra, and many such objects are needed to probe sufficient sight-lines to gain a statistical measure of the IGM, while quasars are apparently exceedingly rare at $z >$ 7.  

What is needed is the ability to derive statistically significant measurements of the neutral fraction at a wide range of redshifts throughout the EoR.  While techniques such as
21 cm line mapping will ultimately provide detailed topographical
information on the neutral state of the IGM, the sensitivity needed
requires the Square Kilometer Array, which will not become fully operational
until into the 2030s.  A technique available \emph{now}, which is
sensitive to a wide range of neutral fractions, probes the entire
epoch in question, and provides information on the spatial nature of
reionization, is the spectroscopic measurement of Ly$\alpha$ emission
from galaxies.
\vspace{-6mm} 

\section{\large Mapping Reionization with Ly$\alpha$ Tomography}
\vspace{-4mm} 

Ly$\alpha$ photons are sensitive probes of reionization as they are resonantly scattered by neutral hydrogen, becoming spatially diffused to extremely low surface brightness levels when passing through regions of neutral H\,{\sc i}, rendering this emission near-impossible to detect.  While this limits the use of Ly$\alpha$ as a spectroscopic redshift tracer for galaxies (especially with 10m telescopes), it makes the observability of this complicated line directly relatable to the neutral fraction in the IGM in the vicinity of an observed galaxy.
In the post-reionization universe Ly$\alpha$ is near ubiquitous, being detectable with 10m telescopes in $\gtrsim$50\% of galaxies at z=6 \citep{stark10}, and is expected to be even more prevalent at higher redshifts, due to a decrease in dust attenuation \citep{stark11,finkelstein12a,bouwens14}.
An observed drop in the fraction of galaxies exhibiting Ly$\alpha$ emission at $z >$ 6 (typically parameterized by comparing the equivalent width [EW] distributions at two redshifts, making significant use of non-detections, modeling all effects of incompleteness) may signal a rapidly evolving IGM neutral fraction.

Due to the recent availability of photometrically-selected $z\!\!=$6--10 galaxy candidates \citep[e.g.,][]{finkelstein15,bouwens15,bowler15,livermore17,salmon18}, and the advent of multi-object red-sensitive optical and near-IR spectrographs (e.g., VLT/FORS2, Keck/MOSFIRE), this experiment is now possible.  Studies have found that Ly$\alpha$ is observable from fewer galaxies than expected at $z\!\!=$6.5--7 ($<$20\%), implying a non-zero neutral fraction in the IGM \citep[e.g.,][]{fontana10,pentericci14,schenker12,ono12,finkelstein13,treu13,tilvi14,stark17,jung18}.  Present constraints are poor due to the low number of detected Ly$\alpha$ emission lines at $z{>}7$.  This is primarily due to the limited spectroscopic sensitivity for such distant galaxies, as low levels of Ly$\alpha$ emission with rest-EW $<$ 20 \AA\ are only accessible for extremely bright galaxies over small fields (m $<$ 26; modestly fainter with lensing, e.g. \citep{hoag17}).  Additionally, as these observations are expensive and even MOSFIRE has a limited field-of-view (FoV; 6$^{\prime}\times4^{\prime}$), current observations probe scales much smaller than the expected sizes of ionized bubbles at the end of reionization (Figure 1).
Consequently, out of the now $>$1000 distant galaxy candidates known, robust (${>}5$-$\sigma$) spectroscopic Ly$\alpha$ detections exist for only a $\sim$dozen galaxies at $z{>}7$.  
While present observations are consistent with a neutral fraction of $\sim$50\% by $z \sim$ 7 \citep{mason18}, the limited spectroscopic depth means that these constraints come predominantly from non-detections, which could occur for a variety of reasons.

Evidence exists that these photons are there to be detected when the sensitivity is high enough.   Recent results from an extremely deep 16 hr integration with MOSFIRE find Ly$\alpha$ detected in modestly bright ($H \sim$ 26) galaxies at $z >$ 7.5 \citep{jung18b}.
Near-IR spectrographs on Extremely Large Telescopes (ELTs; D$>$20 m) will provide a tremendous leap in our spectroscopic sensitivity, allowing the measurement of Ly$\alpha$ emission from faint galaxies for the first time.
Figure 1 shows an image from a reionization simulation highlighting the evolution of the scales of ionized and neutral structures throughout the epoch of reionization, from $\sim$1$^{\prime}$ at $z{\sim}$ 10, to $\sim$20$^{\prime}$ at $z$$\sim$$7$ \citep{ocvirk18}.  This creates a natural synergy between both planned US-led ELTs.  The Giant Magellan Telescope (GMT), with its wide FoV, will be able to probe one or more of these ionized structures at the end of reionization, while the Thirty Meter Telescope (TMT) FoV is well matched to probe the formation of the first reionized regions, where the increased collecting area may be needed.

Ultradeep spectroscopy will also remove a key systematic hindering current analyses: when Ly$\alpha$ is not detected, the spectroscopic redshift of the source is not known.  One must then use the estimate of the photometric redshift probability distribution function to assess the likelihood that the expected wavelength of Ly$\alpha$ is contained in the observed wavelength range.  However, in particular at $z >$ 7, these redshift PDFs are \emph{completely} uncalibrated, and the contamination rate by lower-redshift interlopers is also highly uncertain, and is likely dependent on the quality of the photometric data \citep{pentericci18}.  Beyond Ly$\alpha$, the full rest-frame UV spectra contain UV emission lines from metals, including  
\ciii\ $\lambda\lambda$1907,1909, O\,{\sc iii}] $\lambda\lambda$1661,1666, He\,{\sc ii} $\lambda$1640, and \civ\ $\lambda\lambda$1548, 1550, \citep{stark14,jaskot16}.  
These lines can be comparable in strength to Ly$\alpha$, and thus detectable alongside Ly$\alpha$ in the same observation (if the observed wavelength range is large enough).  This will allow the use of only spectroscopically confirmed galaxies when inferring the neutral fraction, which is not possible today.
These lines will also allow a better understanding of the stellar populations and ISM conditions (see companion white paper by Papovich et al.), enabling more accurate models of the intrinsic Ly$\alpha$ flux from these objects, a key uncertainty in reionization inferences.
\vspace{-6mm} 

\begin{figure*}[t]
\begin{center}
\includegraphics[width=0.95\textwidth]{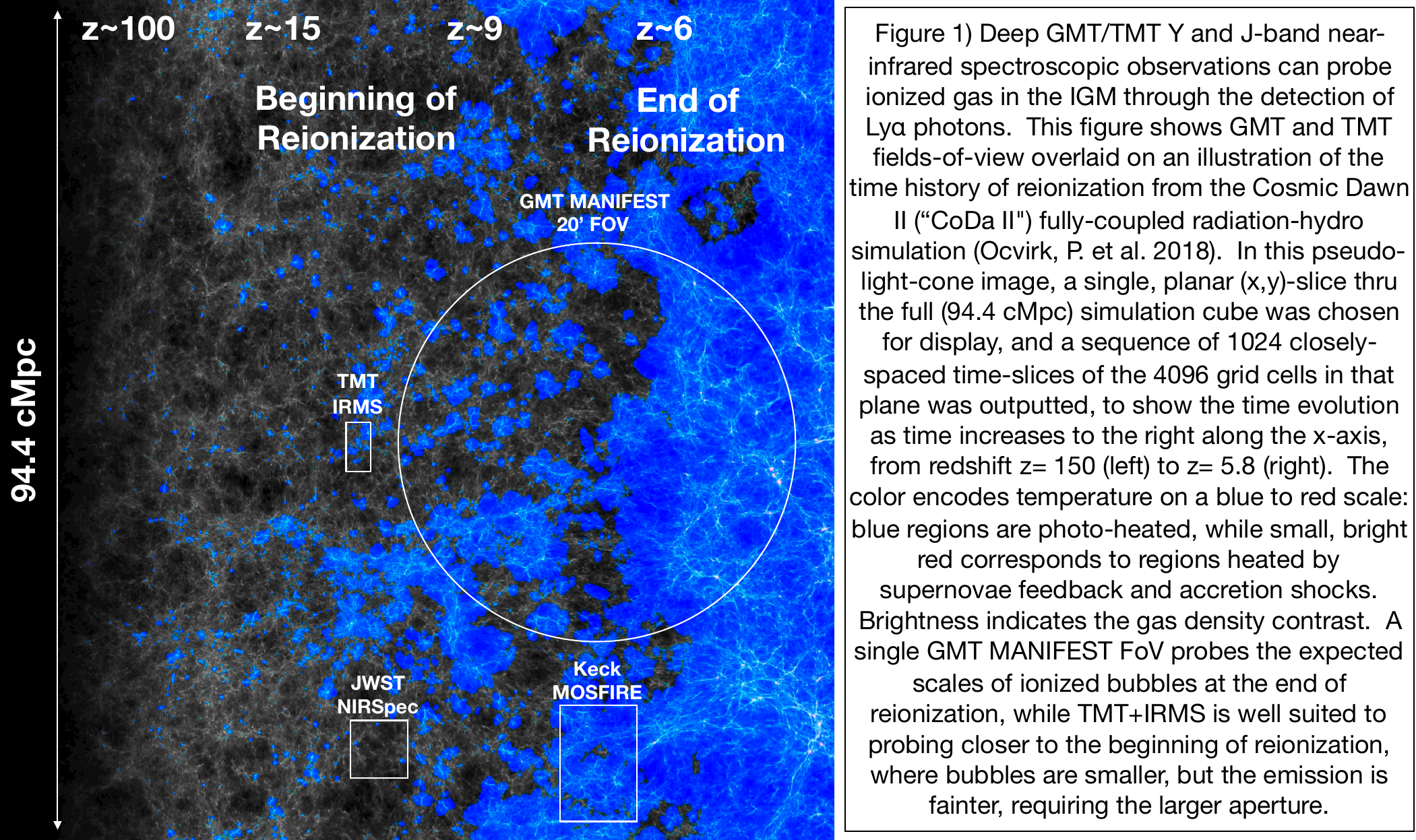}
  \end{center}
  \vspace{-6mm}
  \vspace{-12pt}
\end{figure*}

\section{\large Technological Requirements}
\vspace{-4mm} 

The GMT and TMT change the game by allowing detailed study of faint galaxies in the EoR. Only these facilities will allow the mapping of the evolution of ionized regions in the IGM by measuring Ly$\alpha$ emission over large scales encompassing the expected sizes of ionized bubbles across the EoR.  Here we briefly discuss the technological requirements needed to enable this key science result.
\vspace{1mm}

\noindent \textbf{Sensitivity: }  The key advance here is the large aperture of the GMT/TMT.  Ultra-deep integrations on the Keck 10m telescope still only reach modest Ly$\alpha$ EWs ($\sim$ 30\AA) for bright ($m <$ 26) galaxies \citep{jung18b}.  Fully probing reionization requires reaching EWs $\sim$2$\times$ smaller for galaxies two magnitudes fainter.  In Figure 2, we show an example of a simulated ionized bubble mapped with Ly$\alpha$ tomography. The $\sim$50 Ly$\alpha$ emitters in this region sufficiently sample the extent of the bubble enabling robust estimates of the bubble size and morphology. This population of Ly$\alpha$ emitters was predicted by utilizing the UniverseMachine model \citep{behroozi18} to predict the UV luminosities of halos within the Bolshoi-Planck dark matter simulation \citep{klypin16}. Halos were then assigned intrinsic Ly$\alpha$ EWs and velocity offsets from the \citep{mason18} parameterizations. The intrinsic EWs were then converted into observed EWs by calculating the damping wing transmission fraction \citep{madau00} given the distance from the halo to the neutral boundary along the line of sight and Lyman-alpha velocity offset.  As shown in this figure, the bulk of the galaxies have Ly$\alpha$ EWs of $\sim$15--30 \AA.  For a $m =$ 28 galaxy at $z =$ 8, this corresponds to an integrated emission  line flux of $\sim$5 $\times$ 10$^{-19}$ erg s$^{-1}$ cm$^{-2}$.  Using the exposure time calculator for the GMT concept single-slit SuperFIRE near-IR spectrograph (R.\ Simcoe, private communication), we find that these relatively weak emission lines are detectable in 30 hours at a UV continuum magnitude of 28 (Fig 2, right). For the TMT, the increase in collecting area combined with full AO will allow TMT to detect comparably strong emission lines one magnitude fainter ($m =$ 29) in $\sim$half the exposure time.
\vspace{1mm}

\begin{figure*}[t]
\begin{center}
\hspace{4mm}
\includegraphics[width=0.9\textwidth]{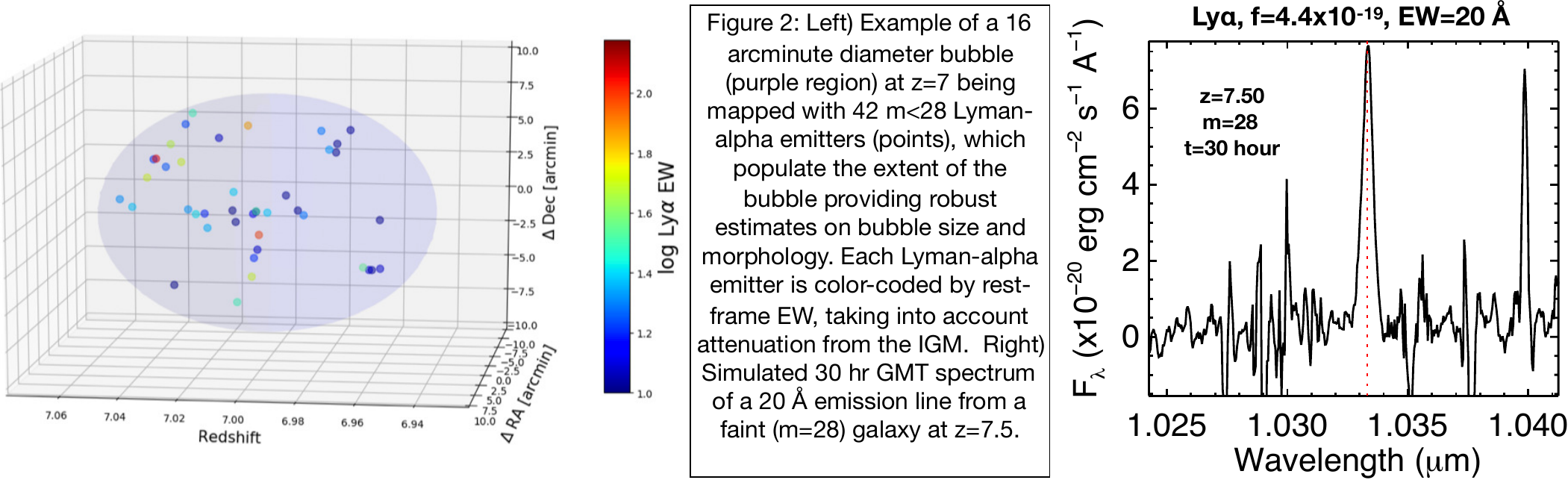}
  \end{center}
  \vspace{-6mm}
  \vspace{-12pt}
\end{figure*}

\noindent \textbf{Instrumentation: }  
The sensitivity achievable by both the GMT and TMT is sufficient for this science goal.  Both telescopes will require multi-object near-IR spectrographs, as ultra-deep exposures will only be worthwhile when done with decently high multiplexing.  With the GMT, the MANIFEST fiber system will patrol a 20$^{\prime}$ diameter field-of-view (where the central $\sim$10$^{\prime}$ benefits from a GLAO correction) with $\sim$1000 0.75$^{\prime\prime}$ fibers.  For this science goal, we require a moderate resolution spectrograph over the $Y$, $J$ and $H$-bands.  The resolution must be moderate (R$\sim$2000-3000) to maximize the space between telluric emission lines, while not splitting the light from these faint, distant galaxies too finely.  A simple instrument consisting of the incident light cross-dispersed into two channels ($Y\!+\!J$ and $H$), each with a HAWAII-4RG detector, fed by a single slit populated by MANIFEST red-side fibers would be suitable.  
Using the reference luminosity functions from \citep{finkelstein16}, we estimate the number of galaxies where Ly$\alpha$ is observable, finding that a single MANIFEST pointing can simultaneously observe $\sim$600 galaxies in the $Y+J$ channel (7 $\lesssim z \lesssim$ 10) to $m=$28, and another $\sim$10 in the $H$ channel (11 $\lesssim z \lesssim$ 14).  A 30 hour ($\sim$4 night) integration with GMT should reach rest-EWs$\sim$15--20 \AA\ at $m=$28, sufficient to map ionized bubbles towards the end of reionization (Fig 2).  Due to the extreme multiplexing advantage, such a spectrograph should be a high priority for early in the GMT's lifetime.

The TMT has a planned near-IR MOS called IRMS, capable of observing up to 46 objects over a 2$^{\prime}$ $\times$ 0.6$^{\prime}$ field-of-view, operating behind the TMT AO system.  
TMT$+$IRMS can best probe towards the beginning of reionization, where the scales of interest are small, encompassed within a single IRMS FoV (Fig 1).  The larger collecting area and full AO correction will enable emission line flux limits near to $\sim$10$^{-19}$ erg s$^{-1}$ cm$^{-2}$ in $\sim$two nights of integration, probing Ly$\alpha$ emission from multiple galaxies $\sim$1 mag deeper than GMT ($m\!\!=$29).

Ly$\alpha$ observations are also possible with {\it JWST}.  While {\it JWST} will measure key physical properties allowing us to estimate the ionizing photon production in galaxies, the {\it JWST}/NIRSpec FoV is very small (Fig 1), thus wide field surveys reaching 1 deg$^2$ are impractical, and the throughput of NIRSPEC drops at $\lambda < $1.2 $\mu$m.  Additionally, the ELTs (with GLAO) have better sensitivity than {\it JWST} in the near-IR.  Finally, though VLT$+$MOONS and Subaru$+$PFS can probe similar spatial scales as GMT's MANIFEST fiber system, the significantly smaller collecting area will make achieving the necessary depths to probe Ly$\alpha$ from fainter galaxies prohibitively expensive.
\vspace{1mm}

\noindent \textbf{The Galaxy Sample:  }
The requirements for this science goal are a large sample of galaxies with relatively high sky density over wide fields.  As shown in Figure 1, the relevant scales at the end of reionization are $\gtrsim$20$^{\prime}$, which sets the minimum region over which we need to select galaxies.  This is larger than any contiguous {\it HST} deep field, and this will also hold true for {\it JWST} (while one $\sim$deg$^2$ survey may happen with Webb, multiple are unlikely given the large overheads).  Performing this study over large contiguous fields, ideally multiple such fields well-separated in the sky, is crucial to overcome both halo cosmic variance and variance in the propagation of ionized bubbles.  By the end of the next decade, when GMT and TMT are operating at full capacity, NASA will have launched the \textit{Wide Field Infrared Survey Telescope} (\textit{WFIRST}).  \textit{WFIRST}'s Wide Field Imager has a FoV of $\sim$0.28 deg$^2$, $\sim$100$\times$ the area of {\it JWST}'s NIRCam.  A single WFIRST pointing allows a contiguous mapping of a field size expected to contain several ionized bubbles to $m <$ 29, and is larger than the GMT$+$MANIFEST FoV (Figure 3).
These observations of Ly$\alpha$ from individual galaxies will be important for determining the purity of whole-sky intensity mapping surveys (e.g. SPHEREx) where $z\sim$1-2 interlopers dominate the $z >$ 7 Ly$\alpha$ intensity mapping signal \citep{pullen14, visbal18}.
\vspace{1mm}

\begin{figure*}[t]
\begin{center}
\hspace{4mm}
\includegraphics[width=0.76\textwidth]{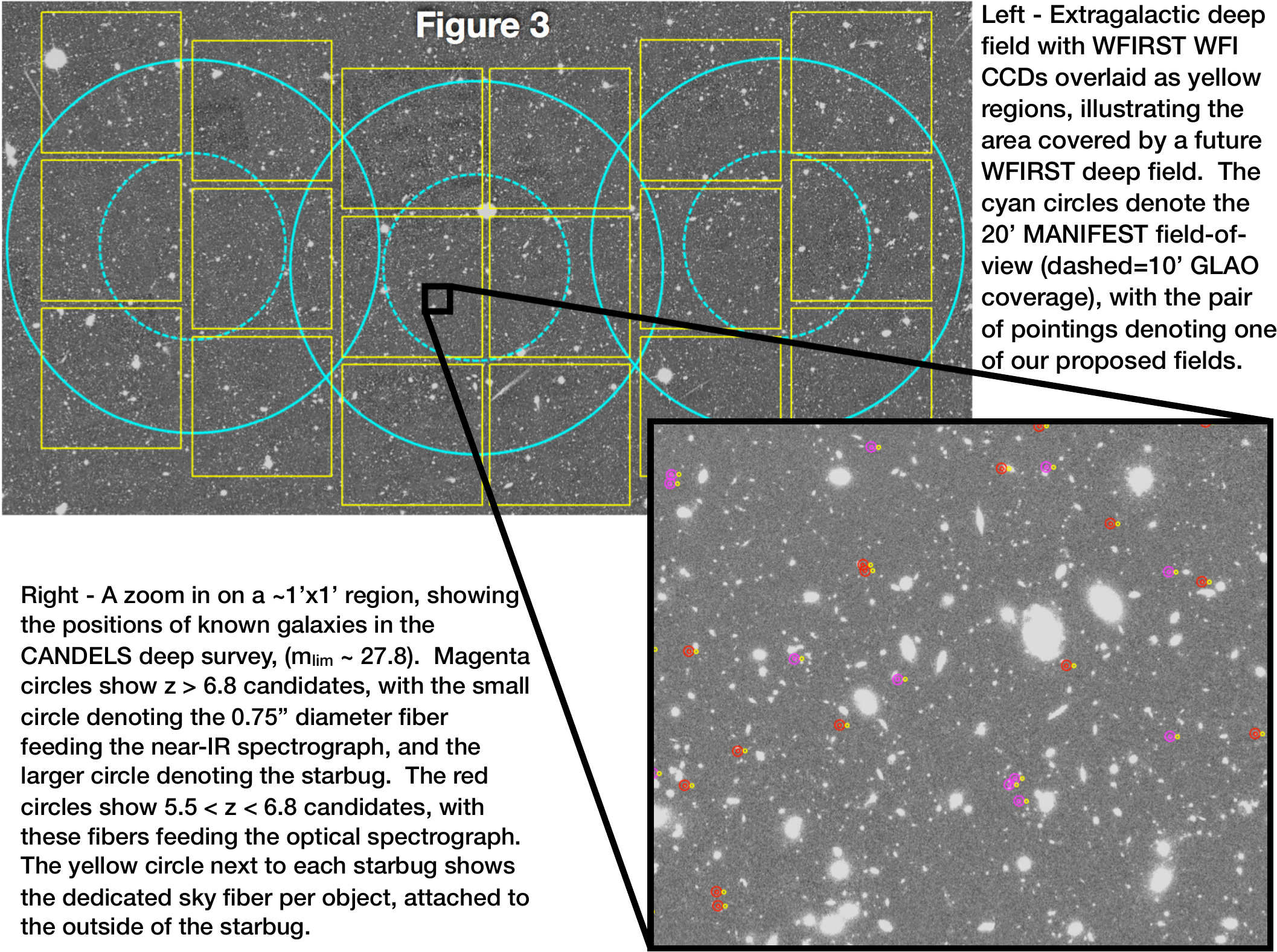}
  \end{center}
  \vspace{-6mm}
  \vspace{-12pt}
\end{figure*}

\noindent \textbf{Summary:  }
Observing Ly$\alpha$ to faint flux levels over  $>$1 deg$^2$ will allow the first direct probes of the IGM ionization state over the scales where we expect to see several ionized and neutral bubbles across the full EoR.  This requires sensitive near-IR spectrographs on 25--30 m-class telescopes (like the GMT and TMT).  These maps of Ly$\alpha$ detections (and limits) will directly trace these ionized bubbles, constraining the neutral fraction by directly measuring the evolution of the Ly$\alpha$ EW distribution (more constraining than the simpler Ly$\alpha$ detection fraction \citep{mason18,jung18}), measuring the correlation function of Ly$\alpha$ emitters, and measuring the cross-correlation between Ly$\alpha$ and 21 cm emission from neutral gas traced by the Square Kilometer Array (SKA; c.\ 2030).  Performing these analyses over several redshift slices from $z=6-10$ will probe the temporal evolution of reionization.  A survey covering the relevant scales and depths during both the early and late stages of the reionization process is possible with a modest investment ($\sim$ a few 10's of nights) on both the GMT and TMT.
Such a survey would provide spectra for $\sim$10,000 galaxies at 6 $\lesssim z \lesssim$ 10, a majority of which we expect to be spectroscopically confirmed via C\,{\sc iii}] and/or Ly$\alpha$ emission.  This will comprise an immense archival sample of galaxy spectra in the EoR, suitable for a variety of analyses, as well as followup observations (e.g., with ALMA, or future space telescopes).
These reionization maps will also provide targets for future, large, space-based telescopes, such as the 15m NASA LUVOIR concept, to perform a full census of galaxies in both ionized and neutral regions, to constrain the impact of reionization on galaxy formation physics.

\pagebreak  
\bibliographystyle{abbrvnat}

\end{document}